\documentclass[aps,prd,onecolumn,groupedaddress,preprintnumbers,superscriptaddress,
showpacs,nofootinbib]{revtex4}
\usepackage{amsmath,amssymb,latexsym,mathrsfs}
\usepackage{color}

\begin{document}

\title{
Angular momentum at null infinity in higher dimensions
}

\author{Kentaro Tanabe}
\affiliation{Yukawa Institute for Theoretical Physics, Kyoto University, Kyoto 606-8502, Japan}

\author{Tetsuya Shiromizu}
\author{Shunichiro Kinoshita}
\affiliation{Department of Physics, Kyoto University, Kyoto 606-8502, Japan}

\preprint{YITP-12-13} 

\begin{abstract}
We define the angular momentum at null infinity in higher dimensions. 
The asymptotic symmetry at null infinity becomes the Poincar\'e group 
in higher dimensions. This fact implies that the angular momentum can 
be defined without any ambiguities such as supertranslation in four 
dimensions. Indeed we can show that the angular momentum in our definition 
is transformed covariantly with respect to the Poincar\'e group. 
\end{abstract}


\maketitle

\section{introduction}

Motivated by the string theory and the scenario with large extra dimensions such 
as the TeV scale gravity \cite{ArkaniHamed:1998rs, Antoniadis:1998ig}, 
the gravitational theory in higher dimensions has been 
investigated \cite{PTP}. Then it has been realized that the higher dimensional gravity 
has much different features from that in four dimensions. 
As one of such differences, there is an issue of the asymptotic structure of the spacetime 
\cite{Hollands:2003ie, Hollands:2003xp, Ishibashi:2007kb, Tanabe:2011es, Tanabe:2009va, Tanabe:2010rm, Tanabe:2009xb}. 
The asymptotically flat spacetime has two 
asymptotic infinities: spatial infinity and null infinity. At spatial infinity we can define the global conserved 
quantities of spacetime such as the mass and angular momentum. In addition 
the multipole moments of spacetime is also defined at spatial infinity 
\cite{Hansen, Tanabe:2010ax}. These multipole moments can be used to classify the 
black hole solutions. At null infinity, the asymptotic structure describes dynamical 
properties of the spacetime because gravitational waves can reach at null infinity. 
Then the study of the asymptotic structure at null infinity is indispensable when one 
considers the dynamical phenomena such as the perturbation 
for black holes and the formation of higher dimensional black holes in 
particle accelerators. As a fundamental aspect of the general relativity, the notion of 
the asymptotic flatness at null infinity is also necessary for the rigorous definition 
of black hole \cite{Hawking:1973uf}. 

The asymptotic structure at null infinity in higher dimensions has been investigated 
using the conformal method in Refs \cite{Hollands:2003ie, Hollands:2003xp, Ishibashi:2007kb}.
In the conformal method, spacetime is conformally embedded into the compact region of 
the another spacetime and the null infinity is defined as the boundary of the spacetime. 
The asymptotic structure at null infinity can be investigated using the introduced  conformal 
factor $\Omega\sim 1/r$ as a coordinate. Therein the null infinity is defined 
on $\Omega=0$. However, there is one problem in this treatment. The gravitational waves 
behave near null infinity with a half integer power of $\Omega$ in odd dimensions. 
At first glance, this shows the non-smoothness of the gravitational fields at null infinity 
in odd dimensions. Because of this non-smooth behavior of the gravitational fields, using the conformal 
method, we cannot define the asymptotic flatness at null infinity in odd dimensions. 

In the Bondi coordinate method Refs. 
\cite{Tanabe:2011es, Bondi:1962px, Sachs:1962wk, Tanabe:2009va, Tanabe:2010rm}, on the other hand, 
we can define the asymptotic flatness at null infinity in arbitrary higher dimensions and safely investigate 
the asymptotic structure at null infinity. In the analysis of the asymptotic structure, it was 
found that the Bondi mass always decreases due to the gravitational waves and the asymptotic 
symmetry at null infinity is the Poincar\'e group. 

It is reminded that the 
asymptotic symmetry is not the Poincar\'e group in four dimensions. 
The asymptotic symmetry in four dimensions is the semi-direct group of the Lorentz group and 
{\it supertranslation}. The supertranslation has the functional degree of freedom and then it is the 
infinite dimensional group. This means that there are infinite directions of the translation which 
causes the ambiguities in the definition of the angular 
momentum. Although there were many efforts to define the angular momentum 
\cite{Prior:1977, Streubel:1978, Winicour:1980, Geroch:1981ut, Dray:1984}, 
there is no sharp definition without any ambiguities in four dimensions.   

In higher dimensions it was shown that the asymptotic symmetry at null infinity is the Poincar\'e group 
\cite{Tanabe:2011es,Tanabe:2009va}. The $n$-dimensional 
Poincar\'e group has the $n$ directions of the translation. 
In this paper, we define the angular momentum 
at null infinity in higher dimensions and shows that it has no 
ambiguities. In fact the angular momentum is transformed covariantly with respect to 
the Poincar\'e group. Note that the study of the angular momentum at null infinity was performed in 
five dimensions \cite{Tanabe:2010rm}. In this paper, we generalize this analysis to arbitrary 
higher dimensions following our previous work of Ref. \cite{Tanabe:2011es}. 

The organization of this paper is as follows. In the next section, we review our previous work on 
null infinity \cite{Tanabe:2011es}. Therein, using the Bondi coordinates, we introduce the definition of
null infinity in arbitrary dimensions. 
We also discussed the asymptotic symmetry at the null infinities 
briefly. In Sec. III, we define the Bondi angular momentum together with the Bondi mass/momentum 
and show its radiation formulae using the 
Einstein equations. In Sec. IV, it will be shown that the angular momentum defined here is 
transformed covariantly under the transformation generated by the asymptotic symmetry. 
Finally we give the summary and outlook in Sec. V.

\section{Review of our previous work}

In this section we review our previous work \cite{Tanabe:2011es}. First we introduce the Bondi coordinates 
adopted here and write down some components of the Einstein equation explicitly. Solving them we specify the 
boundary condition which gives us the definition of the null infinity. Then we discuss the asymptotic symmetries 
at the null infinity. 

\subsection{Bondi coordinates and Einstein equations}

We introduce the Bondi coordinates in $n$ dimensions. First we assume the function $u$ which satisfies 
$\hat\nabla_{a}u \hat\nabla^{a}u =0$ where $\hat\nabla_a$ denotes the
covariant derivative with respect to $n$-dimensional metric $g_{ab}$. 
$u$ is used as the time coordinate. Next the angular coordinate 
$x^{I}$ is defined as $\hat\nabla^{a}u \hat\nabla_{a}x^{I}=g^{uI}=0$. We define the radial coordinate $r$ 
as $\sqrt{\det{g}_{IJ}}=r^{n-2}\omega_{n-2}$ where $\omega_{n-2}$ is the volume element of the unit 
$(n-2)$-dimensional sphere $S^{n-2}$. Then the metric in the Bondi coordinates $x^{a}=(u,r,x^{I})$ 
can be written as
%
\begin{eqnarray}
ds^{2}&=&g_{ab}dx^{a}dx^{b} \notag \\
&=&-Ae^{B}du^{2}-2e^{B}dudr +\gamma_{IJ}(dx^{I}+C^{I}du)(dx^{J}+C^{J}du) \label{Bondi}.
\end{eqnarray} 
%
In this coordinate system, the null infinity is defined at $r=\infty$
and its topology is ${\mathbf R}\times S^{n-2}$. 
For the convenience of our discussion, we define $h_{IJ}$ as $\gamma_{IJ}=r^{2}h_{IJ}$ with 
the following gauge condition
%
\begin{equation}
\sqrt{\det{h_{IJ}}}\,=\,\omega_{n-2}. \label{gauge}
\end{equation} 
%

We provide the Einstein equations in the Bondi coordinates. The vacuum Einstein equations can be 
decomposed into the constraint equation without the $u$ derivative terms and evolution equations 
with the $u$ derivative terms. 

The constraint equations are $\hat{R}_{rr}=0, \hat{R}_{aI}\gamma^{IJ}=0$ and $\hat{R}_{IJ}\gamma^{IJ}=0$. 
$\hat{R}_{ab}$ is the Ricci tensor with respect to 
$g_{ab}$. Using the formulae in Appendix A and Sec. II in Ref. \cite{Tanabe:2011es}, 
we can write the equation $\hat{R}_{rr}=0$ as
%
\begin{equation}
   B' = \frac{r}{4(n-2)}h_{IJ}' h_{KL}' h^{IK}h^{JL} \label{B-eq},
\end{equation}
%
where the prime denotes the $r$ derivative. The equation $\hat{R}_{rJ}\gamma^{IJ}=0$ yields 
%
\begin{equation}
   \frac{1}{r^{n-2}}(r^n e^{-B}h_{IJ}{C^J}')'
    = - {}^{(h)}\nabla_I B' + \frac{n-2}{r}{}^{(h)}\nabla_I B
    + {}^{(h)}\nabla^J h_{IJ}' \label{C-eq},
\end{equation}
%
where ${}^{(h)}\nabla_I$ is the covariant derivative with respect to $h_{IJ}$. 

From $\hat{R}_{IJ}\gamma^{IJ}=0$, we obtain 
%
\begin{eqnarray}
    (n-2)\frac{(r^{n-3}A)'}{r^{n-2}}
    & = &  - {}^{(h)}\nabla_I {C^I}' - \frac{2(n-2)}{r}{}^{(h)}\nabla_I C^I
    - \frac{r^2e^{-B}}{2}h_{IJ}{C^I}'{C^J}' \nonumber \\
    & & - \frac{e^{B}}{2r^2}h^{IJ}{}^{(h)}\nabla_I B {}^{(h)}\nabla_J B
    - \frac{e^{B}}{r^2}{}^{(h)}\nabla_I(h^{IJ}{}^{(h)}\nabla_J B)
    + \frac{e^{B}}{r^2}{}^{(h)}R,
\label{A-eq}
\end{eqnarray}
%
where ${}^{(h)}R$ is the Ricci scalar of $h_{IJ}$. 
Once $h_{IJ}$ is given on a surface $u=u_0$, we can obtain the metric
functions $A, B$ and $C^{I}$ by solving the constraint equations
(\ref{B-eq}), (\ref{C-eq}) and (\ref{A-eq}) on the surface.

The evolution equation is contained in $\hat{R}_{ab}\gamma_I{}^a\gamma_J{}^b=0$ as
%
\begin{eqnarray}
&&e^{-B}\Bigg[ r^{2}\dot{h}_{IJ}^{'}+\frac{n-2}{2}r\dot{h}_{IJ}-\frac{r^{2}}{2}\dot{h}_{IK}h^{'}_{JL}h^{KL}-\frac{r^{2}}{2}\dot{h}_{JK}h^{'}_{IL}h^{KL} \Bigg] \notag \\
&&~~~~~~
-\frac{Ae^{-B}}{2}\Big[ r^{2}h^{''}_{IJ}+(n-2)rh^{'}_{IJ}+2(n-3)h_{IJ}
{- } r^{2}h^{KL}h^{'}_{IK}h^{'}_{JL}\Big]
-\frac{A^{'}e^{-B}}{2}\Big[ r^{2}h^{'}_{IJ} +2rh_{IJ}\Big] \notag \\
&&~~~~~~
-\frac{e^{-B}}{2}\Big[ 2r^{2}\mathcal{L}_{C}h^{'}_{IJ} +(n-2)r\mathcal{L}_{C}h_{IJ}
+r^{2}\mathcal{L}_{C^{'}}h_{IJ}
 - {r^2} h^{'}_{JL}h^{KL}\mathcal{L}_{C}h_{IK} \notag \\
&&~~~~~~
- {r^2} h^{'}_{IL}h^{KL}\mathcal{L}_{C}h_{JK} +{}^{(h)}\nabla_{K}C^{K}
(r^{2}h^{'}_{IJ}+2rh_{IJ}) \Big]  \notag \\
&&~~~~~~
-\frac{e^{-2B}}{2}r^{4}h_{IK}h_{JL}{C^K}'{C^L}' -{}^{(h)}\nabla_{I}{}^{(h)}\nabla_{J}B-\frac{1}{2}{}^{(h)}\nabla_{I}B{}^{(h)}\nabla_{J}B 
+{}^{(h)}R_{IJ}\,=\,0, \label{h-evofull}
\end{eqnarray} 
%
where ${}^{(h)}R_{IJ}$ is the Ricci tensor with respect to $h_{IJ}$ and the dot denotes the $u$ derivative. 
The evolution of $h_{IJ}$ in the Bondi coordinates is determined by Eq.~(\ref{h-evofull}).

\subsection{Asymptotic flatness at null infinity}

The asymptotic flatness at null infinity is defined by the boundary condition at null infinity in the Bondi coordinates (\ref{Bondi}). 
In $n$ dimensions, the boundary conditions for the asymptotic flatness
at null infinity are  
%
\begin{equation}
h_{IJ}\,=\,\omega_{IJ} + O\left(\frac{1}{r^{n/2-1}}\right), \label{h-bound}
\end{equation} 
%
where $\omega_{IJ}$ is the unit round metric on $S^{n-2}$. The boundary conditions for the 
other metric functions are determined by the 
constraint equations of Eqs.~(\ref{B-eq}), (\ref{C-eq}) and (\ref{A-eq}) as \footnote{The condition for 
$B$ will be relaxed to $B=O(r^{-n/2})$ for non-vacuum cases \cite{Godazgar:2012zq}. 
In our present paper, we will focus on the vacuum cases. It is easy to extend our result to non-vacuum cases.}
%
\begin{gather}
A\,=\, 1+ O\left(\frac{1}{r^{n/2-1}}\right), \quad
B\,=\, O\left(\frac{1}{r^{n-2}}\right), \quad
C^{I}\,=\, O\left(\frac{1}{r^{n/2}}\right). \label{ABC-bound}
\end{gather} 
%
Let us see the above solving the constraint equations explicitly. 
We will use some equations for later discussions. First, 
we expand $h_{IJ}$ near null infinity as
%
\begin{equation}
h_{IJ}\,=\,\omega_{IJ} +\sum_{k=0}\frac{h^{(k+1)}_{IJ}}{r^{n/2+k-1}},
\end{equation} 
%
where the summation is taken over $k\in \bf{Z}$ in even dimensions and $2k\in\bf{Z}$ in odd dimensions. 
The indices $I,J,\dots$ are raised and lowered by $\omega_{IJ}$. From the gauge condition of Eq.~(\ref{gauge}), 
we find  that 
$h^{(k+1)}_{IJ}$ is traceless $\omega^{IJ}h^{(k+1)}_{IJ}=0$ for $k<n/2-1$ and, for $k=n/2-1$, 
%
\begin{equation}
\omega^{IJ}h^{(n/2)}_{IJ}\,=\,\frac{1}{2}h^{(1)IJ}h^{(1)}_{IJ}.
\end{equation} 
%

Solving the constraint equation (\ref{B-eq}), we have
%
\begin{equation}
B\,=\,\frac{B^{(1)}}{r^{n-2}} +O\left( \frac{1}{r^{n-3/2}}\right), 
\end{equation} 
%
where
%
\begin{equation}
B^{(1)}\,=\,-\frac{1}{16}\omega^{IK}\omega^{JL}h^{(1)}_{IJ}h^{(1)}_{KL}. 
\end{equation} 
%
From Eq.~(\ref{C-eq}), $C^{I}$ is obtained as
%
\begin{equation}
C^{I}\,=\,\sum_{k=0}^{k<n/2-1}\frac{C^{(k+1)I}}{r^{n/2+k}} +\frac{j^{I}}{r^{n-1}} +O\left( \frac{1}{r^{n-1/2}}\right),
\end{equation} 
%
where, for $k<n/2-1$, 
%
\begin{equation}
C^{(k+1)I}\,=\,\frac{2(n+2k-2)}{(n+2k)(n-2k-2)}\nabla_{J}h^{(k+1)IJ} \label{C-sol}
\end{equation} 
%
and $\nabla_{I}$ is the covariant derivative with respect to $\omega_{IJ}$.
$j^{I}$ is the integration function in the $r$ integration of Eq.~(\ref{C-eq}). 
As seen later, we can see that $j^{I}$ represents the angular momentum of the 
spacetime at null infinity. 

Integrating Eq.~(\ref{A-eq}) we find 
%
\begin{equation}
A\,=\,1+\sum_{k=0}^{k<n/2-2}\frac{A^{(k+1)}}{r^{n/2+k-1}} -\frac{m}{r^{n-3}} +O(r^{-(n-5/2)}), 
\end{equation} 
%
where, for $k<n/2-2$, 
%
\begin{equation}
\begin{aligned}
A^{(k+1)}\,=&\,-\frac{2(n+2k-4)}{(n-2k-4)(n+2k-2)}\nabla^I C_I^{(k+1)}\\
\,=&\,-\frac{4(n+2k-4)}{(n+2k)(n-2k-2)(n-2k-4)}\nabla^{I}\nabla^{J}h^{(k+1)}_{IJ}. 
\end{aligned}
\label{A-sol}
\end{equation} 
%
$m$ is the integration function and reflects the energy-momentum of the spacetime at null infinity. 
For $k=n/2-2$, the left-hand sides of Eqs.~(\ref{C-eq}) and (\ref{A-eq}) vanish. 
Then the right-hand sides of Eqs.~(\ref{C-eq}) and (\ref{A-eq})
provide the following constraint equations
%
\begin{gather}
\nabla_{I}C^{(n/2-1)I}\,=\,0, \\
\nabla^{I}\nabla^{J}h^{(n/2-1)}_{IJ} \,=\,0.  
\end{gather} 
%
In addition, for $k=n/2-1$, we have
\footnote{We found a minor error in Eq. (32) of Ref.~\cite{Tanabe:2011es}, corresponding to Eq. (\ref{B2-sol}) in the current paper. 
But it does not affect the results/all equations presented there except for Eq. (32).} 
%
\begin{equation}
\nabla^{J}h^{(n/2)}_{IJ}\,=\,2\nabla_{I}B^{(1)}
{
+ \nabla_J\left(\frac{1}{2}h^{(1)}_{IK}h^{(1)JK}
	   + \frac{1}{8}\omega_I{}^J h^{(1)}_{KL}h^{(1)KL}
	   \right)
} \label{B2-sol}
. 
\end{equation}
%

We can solve the evolution equation of Eq.~(\ref{h-evofull}) as
%
\begin{eqnarray}
(k+1)\dot{h}^{(k+2)}_{IJ}&=&-\frac{1}{2}\left( n-2k-4 \right) A^{(k+1)}\omega_{IJ} +\frac{1}{8}\Big[n^{2}-6n-(4k^{2}+4k-16) \Big] h^{(k+1)}_{IJ} \notag \\
&&
+\frac{1}{2}\left( -\nabla^{2}h^{(k+1)}_{IJ} +2\nabla_{(I}\nabla^{K}h^{(k+1)}_{J)K} \right) -\frac{1}{2}(n-2k-4)\nabla_{(I}C^{(k+1)}_{J)} 
-\nabla^{K}C^{(k+1)}_{K}\omega_{IJ} \label{h-evo}
\end{eqnarray} 
%
for $k<n/2-1$. The solutions for the higher order of $k\geq n/2-1$ are not important in the following analysis. 
For the convenience of later discussions, we derive the evolution equations for $A^{(k+1)}$ and $C^{(k+1)I}$.
Contracting $\nabla^{J}$ with Eq.~(\ref{h-evo}), we obtain the evolution equation for $C^{(k+1)I}$ as 
%
\begin{eqnarray}
\frac{(k+1)(n+2k+2)}{2(n+2k)}\dot{C}^{(k+2)}_{I}&=& 
-\frac{n-4}{2(n+2k-4)}\nabla_{I}A^{(k+1)} -\frac{1}{4}\nabla^{2}C^{(k+1)}_{I} \notag \\
&&
+\frac{1}{16}\Big[ n^{2}-6n -(4k^{2}+4k-12) \Big]C^{(k+1)}_{I}.  \label{C-evo}
\end{eqnarray} 
%
Contracting $\nabla^{I}$ with Eq.~(\ref{C-evo}) and using the solutions of 
the constraint equations (\ref{C-sol}) and (\ref{A-sol}), 
we have the evolution equation for $A^{(k+1)}$ as 
%
\begin{equation}
\dot{A}^{(k+2)}\,=\,-\frac{n+2k-2}{2(k+1)(n+2k+2)}\nabla^{2}A^{(k+1)}
   +\frac{(n+2k-2)^{2}(n-2k-4)}{8(k+1)(n+2k+2)}A^{(k+1)} 
   \label{A-evo}.
\end{equation} 
%

\subsection{Asymptotic symmetry}
\label{subsec:AS}

The asymptotic symmetry is the global symmetry at null infinity generated by the coordinate 
transformations preserving the gauge and boundary conditions in the Bondi coordinates (\ref{Bondi}). 
The variation of the metric $\delta g_{ab}$ due to the coordinate transformation generated by $\xi^a$ 
is given by 
%
\begin{equation}
\delta g_{ab}\,=\,\hat{\nabla}_{a}\xi_{b}+\hat{\nabla}_{b}\xi_{a},
\end{equation} 
%
where $\hat{\nabla}_{a}$ is the covariant derivative with respect to $g_{ab}$. From Eqs.~(\ref{Bondi}) and 
(\ref{gauge}), the gauge conditions to be satisfied are 
%
\begin{equation}
\delta g_{rr}\,=\,0\,,\quad
\delta g_{rI}\,=\,0\,,\quad
g^{IJ}\delta g_{IJ}\,=\,0. \label{xi-eq}
\end{equation} 
%
From Eqs.~(\ref{h-bound}) and (\ref{ABC-bound}), the boundary conditions to be preserved by the 
coordinate transformations are
%
\begin{equation}
\delta g_{uu}\,=\,O(r^{-(n/2-1)})\,,\quad
\delta g_{uI}\,=\,O(r^{-(n/2-2)})\,,\quad
\delta g_{IJ}\,=\,O(r^{n/2-3}). \label{xi-bound} 
\end{equation} 
%

To satisfy Eq.~(\ref{xi-eq}), the generator of the asymptotic symmetry $\xi$ becomes
%
\begin{gather}
\xi^{u}\,=\, f(u,x^{I}), \\
\xi^{I}\,=\,f^{I}(u,x^{I})+ \int dr\frac{e^{B}}{r^{2}}h^{IJ}\nabla_{J}f ,\\
\xi^{r}\,=\,-\frac{r}{n-2}\left( C^{I}\nabla_{I}f +\nabla_{I}\xi^{I} \right). 
\end{gather} 
%
$f(u,x^{I})$ and $f^{I}(u,x^{I})$ are the integration functions in the $r$ integration 
of the equation $\delta g_{rr}=0$ and $\delta g_{rI}=0$. 
The asymptotic symmetry is the group generated by $f$ and $f^{I}$. 

The boundary conditions (\ref{xi-bound}) give the equations which $f$
and $f^{I}$ should satisfy as 
%
\begin{gather}
\partial_{u}f^{I}\,=\,0 ,\label{fI} \\
\nabla_{I}f_{J}+\nabla_{J}f_{I}\,=\,\frac{2\nabla_K f^K}{n-2}\omega_{IJ},\quad
\nabla_I f^I = (n-2)\frac{\partial f}{\partial u},
\label{fI2} \\
\nabla_{I}\nabla_{J}f\,=\,\frac{\nabla^{2}f}{n-2}\omega_{IJ}\label{f}.
\end{gather}
%
Note that Eq.~(\ref{f}) is required only in $n>4$ dimensions. 
In the following, for the moment, we discuss the asymptotic symmetry in $n>4$ dimensions. 
We will comment on the four dimensional case later. 
 
From Eq.~(\ref{fI}), we find  $f^{I}=f^{I}(x^{I})$. $f^{I}$ is the vector on $S^{n-2}$ and Eq.~(\ref{fI2}) implies that $f^{I}$ generates the conformal isometry
on $S^{n-2}$. The conformal group of $S^{n-2}$ is $\mathrm{SO}(1,n-1)$, which is the Lorentz group. Thus $f^{I}$ is the generator of the Lorentz group.  

Integrating the trace part of Eq.~(\ref{fI2}), we obtain 
%
\begin{eqnarray}
f\,=\, \frac{F(x^{I})}{n-2}u+\alpha (x^{I}), \label{f2}
\end{eqnarray}
%
where $F\,\equiv\,\nabla_{I}f^{I}$ and $\alpha(x^{I})$ is the integration function on $S^{n-2}$. 
Note that the transverse part $f^{\text{(tra)}I}$ of $f^{I}$ which satisfies $\nabla_{I}f^{\text{(tra)}I}=0$ 
is nothing but the Killing vector on $S^{n-2}$, that is, the generator of
$\mathrm{SO}(n-1)$. 
This Killing vector plays an important role in defining the
angular-momentum later.

Now Eq.~(\ref{f}) gives the equation which $\alpha$ should 
satisfy as
%
\begin{eqnarray}
\nabla_{I}\nabla_{J}\alpha \,=\,\frac{1}{n-2}\omega_{IJ} \nabla^{2}\alpha.
\end{eqnarray}
%
The general solutions of this equation are the $l=0$ and $l=1$ modes of the scalar harmonics on $S^{n-2}$. 
Note that the $l=1$ modes satisfy $\nabla_{I}\nabla_{J}\alpha=-\alpha\omega_{IJ}$ too. 
From Eq.~(\ref{f}), we find that $F(x^{I})$ should satisfy $\nabla^{2}F+(n-2)F=0$. The solutions of 
this equation  are the $l=1$ modes of the scalar harmonics on $S^{n-2}$. 
These results mean that the functions $\alpha(x^{I})$ and $F(x^I)$ are 
the generators of the translation and Lorentz boost respectively, and  
$f$ represents the semi-direct property of the Lorentz group and
translation. Then it turns out that the asymptotic symmetry is the semi-direct group of the Lorentz group and translation, 
which is the Poincar\'e group, in $n>4$ dimensions.

In four dimensions, Eqs.~(\ref{fI}) and (\ref{fI2}) are required while Eq.~(\ref{f}) is not. 
Therefore $f^{I}$ generates the Lorentz group and $f$ can be written as Eq.~(\ref{f2}) in $n=4$ dimensions.
However, there are no constraints on $\alpha$ in four dimensions because of the absence of 
Eq.~(\ref{f}). Thus,  
$\alpha(x^{I})$ is the arbitrary function on $S^{2}$
and generates so called supertranslation, not translation. The asymptotic symmetry in four dimensions is 
the semi-direct group of the Lorentz group and the supertranslation. This supertranslation leads the ambiguity 
to the definition of the angular momentum at null infinity in four dimensions.

\section{Bondi angular momentum and radiation formula}

In this section, we will define the Bondi angular momentum. In the pedagogical aspect, we describe the 
definition of the Bondi mass too given in our previous work \cite{Tanabe:2011es}.

\subsection{Bondi mass and angular momentum}

We define the Bondi mass and angular momentum. In the Bondi coordinates, 
$g_{uu}$ and $g_{uI}$ can be expanded near null infinity as
%
\begin{eqnarray}
g_{uu}\,=\,-1-\sum_{k=0}^{k<n/2-2}\frac{A^{(k+1)}}{r^{n/2+k-1}} +\frac{m(u,x^{I})}{r^{n-3}} +O(r^{-(n-5/2)})
\end{eqnarray}
%
and
%
\begin{eqnarray}
g_{uI}\,=\,\sum_{k=0}^{k<n/2-1}\frac{C_{I}^{(k+1)}}{r^{n/2+k-2}} +\frac{j_{I}(u,x^I)+h^{(1)}_{IJ}C^{(1)J}}{r^{n-3}} +O(r^{-(n-5/2)})
\label{guI}.
\end{eqnarray}
%
The functions $m$ and $j_{I}$ are the integration functions and they are free functions 
on the initial surface $u=u_{0}$. 

The Bondi mass $M_{\text{Bondi}}$ and momentum $P^{i}_{\text{Bondi}}$
are defined by \cite{Tanabe:2011es}  
%
\begin{equation}
\begin{aligned}
M_{\text{Bondi}}(u) &\equiv  \frac{n-2}{16\pi}\int_{S^{n-2}}m d\Omega, \\ 
P^{i}_{\text{Bondi}}(u) &\equiv  \frac{n-2}{16\pi}\int_{S^{n-2}}m\hat{x}^{(i)} d\Omega, 
\end{aligned} \label{Bondi-mass}
\end{equation}
%
where $\hat{x}^{(i)}$ is the scalar function on $S^{n-2}$ satisfying 
$\nabla_{I}\nabla_{J}\hat{x}^{(i)}+\omega_{IJ}\hat{x}^{(i)}\,=\,0$. 
These functions are the $l=1$ modes of the scalar harmonic on $S^{n-2}$,
which are defined by $\hat x^{(i)} = x^{(i)}/\rho$ in the Cartesian
coordinates $\{x^{(i)}\}$ of the $(n-1)$-dimensional Euclidean flat space.
Here $S^{n-2}$ is embedded into the $(n-1)$-dimensional Euclidean flat space
as $\rho^2 = \sum_{i=1}^{n-1}(x^{(i)})^2$.
The indices $i$ represent the directions of the translation.
Note that $A^{(k+1)}$ for 
$k<n/2-2$ does not contribute to the global quantities at null infinities because 
$A^{(k+1)}$ and $\hat x^{(i)}A^{(k+1)}$ are written as the form of total derivative 
[see Eq.~(\ref{A-sol})]. For the details, see Eq.~(80) and Appendix B in Ref. 
\cite{Tanabe:2011es}. 

The Bondi energy-momentum vector 
$P^{\mu}_{\text{Bondi}}=(M_{\text{Bondi}}, P^{i}_{\text{Bondi}})$ 
is defined as the $n$-dimensional vector at null infinity. 
In the definition of $P^{\mu}_{\text{Bondi}}$, we introduce the $n$-dimensional vector by 
$\hat{x}^{\mu}=(1,\hat{x}^{(i)})$. This vector can be
naturally identified to the bases in the $n$-dimensional Minkowski
spacetime as mentioned in the next section. Thus the Bondi energy-momentum vector 
$P^{\mu}_{\text{Bondi}}$ can be also regarded as a vector in the Minkowski spacetime. 
In the following, the Greek indices represent the index in Minkowski spacetime.

The Bondi angular momentum $J^{\text{Bondi}}_{(p)}$ is defined by 
%
\begin{equation}
J^{\text{Bondi}}_{(p)}\,=\,-\frac{n-1}{16\pi G}\int_{S^{n-2}}\varphi ^{I}_{(p)}j_{I} d\Omega, 
\label{angular-def}
\end{equation} 
%
where $\varphi^{I}_{(p)}$ is the Killing vector of the round metric 
$\omega_{IJ}$ on $S^{n-2}$. $p$ labels the Killing vectors and 
$1\leq p\leq (n-1)(n-2)/2$. Note that the Bondi angular momentum in 
five dimensions were defined for the Killing vectors which 
commute mutually in Ref. \cite{Tanabe:2010rm}. In this paper, we 
generalized this to define the Bondi angular momentum 
in arbitrary dimensions for all Killing vectors. 
The $\lfloor\frac{n-1}{2}\rfloor$ independent angular momenta are, of course, given by
the mutually commuting Killing vectors.

Here we show that the first term in Eq.~(\ref{guI}) does not contribute to the Bondi 
angular momentum. This is because 
$\varphi^{I}_{(p)}C^{(k+1)}_{I}$ for $k<n/2-1$ can be written as the total derivative as
%
\begin{eqnarray}
\varphi^{I}_{(p)}C^{(k+1)}_{I}&=&\frac{2(n+2k-2)}{(n+2k)(n-2k-2)}\varphi^{I}_{(p)}\nabla^{J}h^{(k+1)}_{IJ} \notag \\
&=&\frac{2(n+2k-2)}{(n+2k)(n-2k-2)}\nabla^{J}\left( \varphi^{I}_{(p)}h^{(k+1)}_{IJ} \right),
\end{eqnarray} 
%
where we used Eq.~(\ref{C-sol}) and the Killing equation $\nabla_{I}\varphi_{(p)J}+\nabla_{J}\varphi_{(p)I}=0$. 
The term of $h_{IJ}^{(1)}C^{(1)J}$ in Eq.~(\ref{guI}) is nothing, but it just comes from 
the lowering of the index of the metric. Therefore, we will not think that it contributes to the 
angular momentum. This will be also confirmed later when one considers the transformation property generated 
by asymptotic symmetry at null infinity (Sec.~\ref{sec:covariance}).

\subsection{Radiation formula}

The functions $m$ and $j_{I}$ are free functions on the initial surface $u=u_{0}$. 
The evolutions of these quantities are determined
from the Einstein equations. The Einstein equation $\hat{R}^{rr}=0$ (see Eq.~(16) in Ref. 
\cite{Tanabe:2011es}) can be expanded near null infinity as
%
\begin{equation}
\hat{R}^{rr}\,=\,\sum_{k=0}^{k<n/2-2}\frac{(\hat{R}^{rr})^{(k+1)}}{r^{n/2+k-1}} +\frac{(\hat{R}^{rr})^{(n/2-1)}}{r^{n-3}} +O\left( \frac{1}{r^{n-5/2}}\right). 
\end{equation} 
%
The equations $(\hat{R}^{rr})^{(k+1)}=0$ for $k<n/2-2$ provide us Eq.~(\ref{A-evo}) again 
and has no new informations. This feature is guaranteed by the Bianchi identity. 
The equation $(\hat{R}^{rr})^{(n/2-1)}=0$
describes the evolution of the function $m$ as
%
\begin{eqnarray}
   \dot{m}\,=\,-\frac{1}{2(n-2)}\dot{h}_{IJ}^{(1)}\dot{h}^{(1)IJ}
   +\frac{n-5}{n-2}\nabla^{I}C^{(n/2-2)}_{I}
   +\frac{1}{n-2}\nabla^{2}A^{(n/2-2)}.
\end{eqnarray}
%
Integrating this equation on the unit $(n-2)$-dimensional sphere, we obtain the Bondi mass-loss law
%
\begin{eqnarray}
   \frac{d}{du}M_{\text{Bondi}}\,=\,-\frac{1}{32\pi}\int_{S^{n-2}}
   \dot{h}_{IJ}^{(1)}\dot{h}^{(1)IJ} d\Omega \leq 0 
   \label{BML}.
\end{eqnarray}
%
The above implies that the Bondi mass always 
decreases by radiating the gravitational waves in any dimensions. In other words, 
the gravitational waves carry the positive energy flux to null infinity in any dimensions. 

The Einstein equations $\hat{R}^{rI}=0$ contain the evolution equations of $j_{I}$. $\hat{R}^{rI}$ 
can be expanded near null infinity as (see Eq.~(19) in Ref. \cite{Tanabe:2011es})
%
\begin{equation}
\varphi_{(p)I}\hat{R}^{rI}\,=\,\sum_{k=0}^{k<n/2-1}
\frac{\varphi_{(p)I}(\hat{R}^{rI})^{(k+1)}}{r^{n/2+k-1}}
 +\frac{\varphi_{(p)I}(\hat{R}^{rI})^{(n/2)}}{r^{n-2}} 
+O\left( \frac{1}{r^{n-3/2}}\right). 
\end{equation} 
%
The equations $\varphi_{(p)I}(\hat{R}^{rI})^{(k+1)}=0$ for $k<n/2-1$ provide us Eq.~(\ref{C-evo}) again. 
It is also guaranteed by the Bianchi identity. 
The equation $\varphi_{(p)I}(\hat{R}^{rI})^{(n/2)}=0$ presents the evolution equation of $j_{I}$ as
%
\begin{eqnarray}
-(n-1)\varphi^{I}_{(p)}\dot{j}_{I}&=&\varphi^{I}_{(p)} \Bigg[ \partial_{u}(h^{(1)}_{IJ}\nabla_{K}h^{(1)JK}) +h^{(1)JK}\nabla_{J}\dot{h}^{(1)}_{IK} 
+\nabla_{K}h^{(1)JK}\dot{h}^{(1)}_{IJ} +\frac{1}{2}\dot{h}^{(1)JK}\nabla_{I}h^{(1)}_{JK}\Bigg] \notag \\
&&
-\varphi^{I}_{(p)}\Big[ \nabla_{I}m {-}
\nabla_{I}\dot{B}^{(1)}
{+} \nabla^{J}\dot{h}^{(n/2)}_{IJ}
{+} (n-3)\nabla^{J}h^{(n/2-1)}_{IJ} \Big] 
+2\varphi^{I}_{(p)}\nabla^{J}\nabla_{(I}C^{(n/2-1)}_{J)} \notag \\
&=&\varphi^{I}_{(p)} \Bigg[ 2\dot{h}^{(1)}_{IJ}\nabla_{K}h^{(1)JK}
-\nabla^{K}h^{(1)IJ}\dot{h}^{(1)}_{JK} +\frac{1}{2}\dot{h}^{(1)JK}\nabla_{I}h^{(1)}_{JK}\Bigg] \notag \\
&&+
\nabla_{I}\Bigg[ \varphi^{J}_{(p)}\partial_{u}(h^{(1)}_{JK}h^{(1)IK}) +\varphi^{I}_{(p)}(-m +\dot{B}^{(1)}) -\varphi_{(p)J}\dot{h}^{(n/2)IJ} \notag \\
&&
-(n-3)\varphi_{(p)J}h^{(n/2-1)IJ} + \varphi^{J}_{(p)}\nabla^{I}C^{(n/2-1)}_{J}-C^{(n/2-1)}_{J}\nabla^{I}\varphi^{J}_{(p)} \Bigg], \label{j-radiation}
\end{eqnarray} 
%
where we used the Killing equation, $\nabla_{I}\varphi_{(p)J}+\nabla_{J}\varphi_{(p)I}=0$. 

Then, we can obtain the radiation formula of the Bondi angular momentum $J^{\text{Bondi}}_{(p)}$ as
%
\begin{eqnarray}
\frac{d}{du}J^{\text{Bondi}}_{(p)}\,=\,\frac{1}{16\pi G}\int_{S^{n-2}}\varphi^{I}_{(p)} \Bigg[ 2\dot{h}^{(1)}_{IJ}\nabla_{K}h^{(1)JK}
-\nabla_{K}h_{IJ}^{(1)}\dot{h}^{(1)JK} +\frac{1}{2}\dot{h}^{(1)JK}\nabla_{I}h^{(1)}_{JK}\Bigg] d\Omega. \label{angular-radiation}
\end{eqnarray}
%
This equation shows that the Bondi angular momentum is changed when the spacetime has time and angular dependences. 
The radiation formula (\ref{angular-radiation}) is natural in this sense.

\section{Poincar\'e covariance}
\label{sec:covariance}

In this section we consider the transformation of our Bondi mass $M_{\text{Bondi}}$ and angular momentum
$J^{\text{Bondi}}_{(p)}$ generated by the asymptotic symmetry. The validity of our definitions of 
the Bondi mass and angular momentum will be supported by the fact that the Bondi mass and angular 
momentum are transformed covariantly with respect to the asymptotic symmetry.

\subsection{Poincar\'e covariance}

Let us investigate the transformation rule of the Bondi energy-momentum $P^{\mu}_{\text{Bondi}}$ and 
angular momentum $J^{\text{Bondi}}_{(p)}$ by the asymptotic symmetry. In particular, we focus the cases with $f=\alpha$ and $f^{I}=0$, 
which is the translation of the Poincar\'e group in $n>4$ dimensions.

As we mentioned in Sec.~\ref{subsec:AS}, $\alpha(x^I)$ can be decomposed
into the $l=0$ and $l=1$ modes of the scalar harmonics on $S^{n-2}$.
Using these harmonics as bases $\hat{x}^{\mu}=(1,\hat{x}^{(i)})$, 
we can naturally introduce the translational vector $\alpha_{\mu}$ in the $n$-dimensional Minkowski spacetime defined by 
$\alpha(x^I) = \alpha_{\mu}\hat{x}^{\mu}$. 
Moreover, because the asymptotic symmetry at null infinity is the 
Poincar\'e group, we can identify asymptotic structure at null 
infinity with the $n$-dimensional Minkowski spacetime and then 
obtain a natural map between quantities at null infinity and 
those of vector spaces in the Minkowski spacetime. 
Then we can discuss the transformations of $P^{\mu}_{\text{Bondi}}$ 
and $J^{\text{Bondi}}_{(p)}$ by the translational vector $\alpha_{\mu}$ in the Minkowski spacetime.

In general, the energy-momentum vector $P_{\mu}$ and angular momentum $M_{\mu\nu}$
in the Minkowski spacetime are expected to be transformed by 
translation of the Poincar\'e group as
%
\begin{equation}
\begin{aligned}
P_{\mu}&\rightarrow P_{\mu}, \\
M_{\mu\nu}&\rightarrow M_{\mu\nu} -2P_{[\mu}\alpha_{\nu]},
\end{aligned} \label{poincare}
\end{equation}
%
where $\alpha_\mu$ is a translational vector.
However, since the gravitational waves carry the energy and angular momentum to null infinity, 
the Bondi energy-momentum $P^{\mu}_{\text{Bondi}}$ and angular momentum 
$J^{\text{Bondi}}_{(p)}$ are changed under the translation. Then, taking these 
effects into account, $P^{\mu}_{\text{Bondi}}$ and $J^{\text{Bondi}}_{(p)}$
should be transformed as
%
\begin{equation}
\begin{aligned}
P_{\text{Bondi}}^{\mu}&\rightarrow P_{\text{Bondi}}^{\mu}
 + \alpha_\nu \frac{d}{du}P^{\mu\nu}_{\text{Bondi}}, \\
M^{\text{Bondi}}_{\mu\nu}&\rightarrow M^{\text{Bondi}}_{\mu\nu}
 - 2P^{\text{Bondi}}_{[\mu}\alpha_{\nu]}
 + \alpha^\rho\frac{d}{du}M^{\text{Bondi}}_{\mu\nu\rho},
\end{aligned} \label{Bondi-poincare}
\end{equation}
%
instead of Eq.~(\ref{poincare}). 
Note that the each space-space component of $M^{\text{Bondi}}_{\mu\nu}$
corresponds to $J^{\text{Bondi}}_{(p)}$.
From now on, we will confirm these equations. The last terms in each transformations come from 
the effect of radiations and the concrete expressions will be given later. 

The generator of the translation $f=\alpha$ and $f^{I}=0$ can be expanded near null infinity as
%
\begin{eqnarray}
\xi^{u}&=&\alpha(x^{I}), \\
\xi^{I}&=&-\frac{1}{r}\nabla^{I}\alpha +\sum_{k=0}^{k<n/2-1}\frac{2h^{(k+1)IJ}\nabla_{J}\alpha}{n+2k} \frac{1}{r^{n/2+k}}  \notag \\
&&
-\frac{1}{n-1}\frac{1}{r^{n-1}}\left( B^{(1)}\nabla^{I}\alpha
 -h^{(n/2)IJ}\nabla_{J}\alpha
 {+ } h^{(1)IL}h^{(1)J}_{L}\nabla_{J}\alpha \right)
 +O(r^{-(n-1/2)}) ,
\\
\xi^{r}&=&\frac{\nabla^{2}\alpha}{n-2} - 
\sum_{k=0}^{k<n/2-1}\frac{2}{n+2k-2}
\frac{ 
{C^{(k+1)I}\nabla_{I}\alpha} 
}{r^{n/2+k-1}}
 +O(r^{-(n-2)}).
\end{eqnarray}
%

\subsection{Covariance of Bondi energy-momentum }

Following Ref. \cite{Tanabe:2011es}, we briefly sketch the argument to show the covariance of the 
Bondi energy-momentum. 
The Bondi energy-momentum $P^{\mu}_{\text{Bondi}}$ is defined from $g_{uu}$ as in Eq.~(\ref{Bondi-mass}). 
To find the variation of $m$, we 
look at the variation $\delta g_{uu}$. $\delta g_{uu}$ can be expanded near null infinity as
%
\begin{eqnarray}
 \delta g_{uu}&=& 2\hat{\nabla}_{u}\xi_{u} \notag \\
 &=& \sum_{k=0}^{k<n/2-2}\delta g^{(k+1)}_{uu}r^{-(n/2+k-1)} +\frac{\delta m}{r^{n-3}} +O(r^{-(n-5/2)}),  
\end{eqnarray}
%
where
%
\begin{eqnarray}
   \delta g^{(k+1)}_{uu}&=&\frac{2}{n+2k}[\nabla^{2}(\alpha A^{(k)})+(n-2)\alpha A^{(k)}]
   +\frac{4}{(n+2k)(n-2k-2)}\nabla^{I}\nabla^{J}(\alpha\dot{h}^{(k+1)}_{IJ})\notag \\
   &&-\frac{2(n+2k-6)}{(n+2k)(n-2k-2)}[\nabla^{I}\nabla^{J}(\nabla_{I}\alpha C^{(k)}_{J})
+C^{(k)}_{I}\nabla^{I}\alpha ] \label{A-trans}
\end{eqnarray}
%
for $0\leq k<n/2-2 $. $\delta m$ is given by
%
\begin{eqnarray}
    \delta m &=&\,\alpha\dot{m}+\frac{2}{n-3}\nabla^{I}\alpha\dot{C}_{I}^{(n/2-1)}
    -(n-4)\alpha A^{(n/2-2)}+\nabla^{I}\alpha \nabla_{I}A^{(n/2-2)} \notag \\ 
    &=&-\frac{\alpha}{2(n-2)}\dot{h}^{(1)}_{IJ}\dot{h}^{(1)IJ}
    +\frac{1}{n-2}\Big[ \nabla^{2}(\alpha A^{(n/2-2)})+(n-2)\alpha 
    A^{(n/2-2)} \Big] \notag \\
&&-\frac{n-5}{n-2}\Big[ \nabla^{I}\nabla^{J}(\nabla_{I}\alpha C_{J}^{(n/2-2)})
    +C^{(n/2-2)}_{I}\nabla^{I}\alpha \Big], \label{m-trans}
   \end{eqnarray}
%
where we used Eqs.~(\ref{A-sol}) and (\ref{A-evo}). 
From Eq.~(\ref{A-trans}) and the fact that $\hat{x}^{(i)}$ satisfies 
$\nabla_{I}\nabla_{J}\hat{x}^{(i)}+\hat{x}^{(i)}\omega_{IJ}=0$, we see  
%
\begin{gather}
\int_{S^{n-2}}\delta g_{uu}^{(k+1)}\,=\,0
\end{gather}
and
\begin{gather}
\int_{S^{n-2}}\hat{x}^{(i)}\delta g_{uu}^{(k+1)}\,=\,0
\end{gather} 
%
for $k<n/2-2$. This means that $g_{uu}^{(k+1)}$ for $k<n/2-2$ does not contribute to the global quantities 
in the transformed Bondi coordinates. 

The variation of the Bondi energy-momentum $\delta P^{\mu}_{\text{Bondi}}$ can be obtained 
by integrating Eq.~(\ref{m-trans}) as 
%
\begin{eqnarray}
\delta P^{\mu}_{\text{Bondi}}&=& \frac{n-2}{16\pi G}\int_{S^{n-2}} \hat{x}^{\mu}\delta m d\Omega  \notag\\
&=& -\frac{1}{16\pi G} \int_{S^{n-2}}
 \alpha\hat{x}^{\mu}\dot{h}^{(1)}_{IJ}\dot{h}^{(1)IJ} d\Omega  \notag\\
&=& - \alpha_\nu \frac{1}{16\pi G} \int_{S^{n-2}}
 \hat{x}^{\mu}\hat{x}^{\nu}\dot{h}^{(1)}_{IJ}\dot{h}^{(1)IJ} d\Omega . 
\end{eqnarray} 
%
In the last line of the above, we used $\alpha=\alpha_\mu \hat x^\mu$. 
Then we regard the right-hand side of this equation as 
$\alpha_\nu dP^{\mu\nu}_{\text{Bondi}}/du$ in Eq.~(\ref{Bondi-poincare})
%
\begin{eqnarray}
\frac{d}{du}P^{\mu \nu}_{\text{Bondi}}=- \frac{1}{16\pi G} \int_{S^{n-2}}
 \hat{x}^{\mu}\hat{x}^{\nu}\dot{h}^{(1)}_{IJ}\dot{h}^{(1)IJ} d\Omega . 
\end{eqnarray} 
%
This means that 
the Bondi energy-momentum defined is transformed covariantly with respect to the Poincar\'e group. 
In particular, since the time-component becomes 
$dP^{\mu 0}_{\text{Bondi}}/du = dP^{\mu}_{\text{Bondi}}/du$, 
we have 
\begin{equation}
 P^{\mu}_{\text{Bondi}} \rightarrow P_{\text{Bondi}}^{\mu}
  + \alpha \frac{d}{du}P_{\text{Bondi}}^{\mu},
\end{equation}
for the time-translation.

\subsection{Poincar\'e covariance of Bondi angular momentum}

Next  we investigate the variation of the Bondi angular momentum $J^{\text{Bondi}}_{(p)}$ 
by the translation. The Bondi angular momentum $J^{\text{Bondi}}_{(p)}$
is identified with a space-space component 
of $M^{\text{Bondi}}_{\mu\nu}$ in Eq.~(\ref{Bondi-poincare}). Thus, in the following, we consider the space-space components.
Note that the time-space component of $M^{\text{Bondi}}_{\mu\nu}$
represents the Lorentz boost.
The $l=0$ mode of $\alpha$ generates the time-translation and the $l=1$
modes generate the translations in the spatial directions.

The Bondi angular momentum is defined using a part of 
$g_{uI}$ as Eq. (\ref{angular-def}). The variation of the Bondi angular momentum is given by 
%
\begin{eqnarray}
\delta J_{(p)}^{\text{Bondi}}=-\frac{n-1}{16\pi G}\int_{S^{n-2}}\varphi^{I}_{(p)}\delta j_{I} d\Omega. 
\end{eqnarray} 
%

The variation $\delta g_{uI}$ can be expanded as
%
\begin{eqnarray}
\delta g_{uI}&=& \hat{\nabla}_{u}\xi_{I} +\hat{\nabla}_{I}\xi_{u} \notag \\
&=:& \sum_{k=0}^{k<n/2-1}\frac{\delta g^{(k+1)}_{uI}}{r^{n/2+k-2}}
 + \frac{\delta g^{(n/2)}_{uI}}{r^{n-3}} + O(r^{-(n-5/2)}),
\end{eqnarray} 
%
where 
%
\begin{eqnarray}
\delta g _{uI}^{(k+1)}&=& 
-A^{(k)}\nabla_{I}\alpha
 + \frac{2}{n+2k-4}\nabla_{I}(C^{(k)J}\nabla_{J}\alpha) 
-C^{(k)J}\nabla_{I}\nabla_{J}\alpha
-\nabla^{J}\alpha \nabla_{J}C^{(k)}_{I}
\notag \\ 
&&-\frac{n+2k-6}{2(n-2)}C^{(k)}_{I}\nabla^{2}\alpha
+ \alpha \dot{C}_I^{(k+1)}
+ \frac{2}{n+2k}\dot{h}^{(k+1)}_{IJ}\nabla^{J}\alpha
\label{delta_guI}
\end{eqnarray}
%
for $k<n/2-1$.
Then we find 
%
\begin{eqnarray}
\varphi^{I}_{(p)} \delta g _{uI}^{(k+1)}
&=& 
\nabla_{I}\Bigg[
\frac{2}{n+2k-4}\varphi^{I}_{(p)}C^{(k)J}\nabla_{J}\alpha
 -\frac{4(n-4)}{(n+2k)(n+2k-6)}\varphi^{I}_{(p)}\alpha A^{(k)} \notag \\
&&~~
+\frac{2\Big[ n^{2}+(4k-10)n +(4k^{2}-12k+16) \Big]}{(n+2k)(n-2k-2)(n+2k-4)}C^{(k)I}\varphi^{J}_{(p)}\nabla_{J}\alpha \notag \\
&&~~
-\frac{2}{n+2k}\Big[ \alpha \varphi_{(p)J}\nabla^{I}C^{(k)J}
 -\varphi^{J}_{(p)}C^{(k)}_{J}\nabla^{I}\alpha -\alpha C^{(k)}_{J} 
\nabla^{I}\varphi^{J}_{(p)} \Big] 
 - \varphi_{(p)}^J C_J^{(k)}\nabla^I\alpha \notag \\
&&~~
+ \frac{n+2k-4}{n+2k} \alpha \varphi_{(p)J}h^{(k)IJ} 
+ \frac{2(n+2k-6)}{(n+2k)(n-2k-2)}
h^{(k)IJ}\nabla^{K}\alpha \nabla_{K}\varphi_{(p)J}
\notag \\
&&~~
+ \frac{n^2 - 8n - 4k^2 + 20}{(n-2)(n+2k)(n-2k-2)}
\varphi_{(p)J}h^{(k)IJ} \nabla^2 \alpha \Bigg] \label{guI-trans},
\end{eqnarray}
%
where we used Eqs.~(\ref{C-sol}), (\ref{A-sol}) and (\ref{C-evo}).
Also we used the Killing equation 
$\nabla_{I}\varphi_{(p)J}+\nabla_{J}\varphi_{(p)I}=0$ 
and $\nabla_{I}\nabla_{J}\alpha=\omega_{IJ}\nabla^2\alpha/(n-2)$. 
Thus we could confirm again that 
$g_{uI}^{(k+1)}$ for $k<n/2-1$ does not contribute to the global quantities 
because it can be written by the total derivative as Eq.~(\ref{guI-trans}). 
For $k=n/2-1$, on the other hand, the variation becomes
\begin{eqnarray}
\delta g_{uI}^{(n/2)} &=& \delta (j_{I}+h^{(1)}_{IJ}C^{(1)J}) \notag \\
&=&
\alpha \partial_{u}(j_{I}+h^{(1)}_{IJ}C^{(1)J}) +\frac{2}{n}h^{(1)}_{IJ}\partial_{u}h^{(1)JL}\nabla_{L}\alpha 
+ { \frac{1}{n-3} }
\nabla_{I}(C^{(n/2-1)}_{J}\nabla^{J}\alpha) \notag \\
&&+m\nabla_{I}\alpha -C^{(n/2-1)J}\nabla_{I}\nabla_{J}\alpha 
-\frac{1}{n-1}\left( \partial_{u}B^{(1)} \nabla_{I}\alpha
	       -\dot{h}^{(n/2)}_{IJ}\nabla^{J}\alpha
 {+ }  
\partial_{u}
(h^{(1)}_{IJ}h^{(1)JK})\nabla_{K}\alpha\right) \notag \\
&&-\frac{n-4}{n-2}C^{(n/2-1)}_{I} \nabla^{2}\alpha -\nabla^{J}\alpha \nabla_{J}C^{(n/2-1)}_{I} . \label{jhC-trans}
\end{eqnarray}
%
We must evaluate $\delta j_I$ to see the variation of the Bondi angular momentum. 
Therefore, we should subtract the variation
$\delta (h^{(1)}_{IJ}C^{(1)J})$ from Eq.~(\ref{jhC-trans}). 
The variation $\delta C^{(1)I}$ is given by 
%
\begin{eqnarray}
\delta C^{(1)I}\,=\,\frac{2}{n}\nabla_{J}\alpha \dot{h}^{(1)IJ} +\alpha\dot{C}^{(1)I}, \label{C-trans}  
\end{eqnarray}
%
from Eq.~(\ref{delta_guI}) for $k=0$.
Since $\delta g_{IJ}$ is 
%
\begin{eqnarray}
\delta g_{IJ} &=& \hat{\nabla}_{I}\xi_{J} + \hat{\nabla}_{J}\xi_{I} \notag \\
&=&r^{2}\left(\frac{\alpha\dot{h}^{(1)}_{IJ}}{r^{n/2-1}} +O(r^{-(n/2-1/2)})  \right), 
\end{eqnarray} 
%
we find $\delta h^{(1)}_{IJ}=\alpha\dot{h}^{(1)}_{IJ}$.
Then we have 
%
\begin{equation}
\delta (h^{(1)}_{IJ}C^{(1)J})\,=\,
{
\alpha\left(\dot{h}^{(1)}_{IJ}C^{(1)J} +
       h^{(1)}_{IJ}\dot{C}^{(1)J}\right)
+ \frac{2}{n}h^{(1)}_{IJ} \dot{h}^{(1)JK} \nabla_{K}\alpha.
}
\end{equation}
%
Subtracting the above from Eq.~(\ref{jhC-trans}), we obtain 
%
\begin{eqnarray}
\varphi^{I}_{(p)}\delta j_{I}&=& \varphi^{I}_{(p)}\Bigg[\alpha \dot{j}_{I} 
+\frac{1}{n-3}\nabla_{I}(C^{(n/2-1)}_{J}\nabla^{J}\alpha) 
+m\nabla_{I}\alpha -C^{(n/2-1)J}\nabla_{I}\nabla_{J}\alpha \notag \\
&&~~
-\frac{1}{n-1}\left( \dot{B}^{(1)} \nabla_{I}\alpha
	       -\dot{h}^{(n/2)}_{IJ}\nabla^{J}\alpha + \partial_{u}
(h^{(1)}_{IJ}h^{(1)JK})\nabla_{K}\alpha\right) \notag \\
&&~~
 -\frac{n-4}{n-2}C^{(n/2-1)}_{I} \nabla^{2}\alpha -\nabla^{J}\alpha \nabla_{J}C^{(n/2-1)}_{I} \Bigg] \notag \\
 &=&
-\frac{\alpha}{n-1}\varphi^{I}_{(p)} \Bigg[ 2\dot{h}^{(1)}_{IJ}\nabla_{K}h^{(1)JK}
-\nabla_{K}h_{IJ}^{(1)}\dot{h}^{(1)JK} +\frac{1}{2}\dot{h}^{(1)JK}\nabla_{I}h^{(1)}_{JK}\Bigg] +\frac{n-2}{n-1}m\varphi^{I}_{(p)}\nabla_{I}\alpha\notag \\
&&+\nabla_{I}\Bigg[  -\frac{1}{n-1}\Big[ \alpha \varphi^{I}_{(p)}(\dot{B}^{(1)}-m) 
-\alpha \varphi_{(p)J}\dot{h}^{(n/2)IJ}
+ \partial_{u}(h^{(1)}_{IJ}h^{(1)JK})\nabla_{K}\alpha
\notag \\
&&
 + \alpha \varphi^{J}_{(p)}\nabla^{I}C^{(n/2-1)}_{J}
 - \alpha C^{(n/2-1)}_{J}\nabla^{I}\varphi^{J}_{(p)} \Big] 
 + \frac{1}{n-3}\varphi^{I}_{(p)}C^{(n/2-1)J}\nabla_{J}\alpha
 - \frac{n-2}{n-1}\varphi^{J}C^{(n/2-1)}_{J}\nabla^{I}\alpha  \notag \\
&&
 + \frac{n-3}{n-1} 
 \Big[\varphi_{(p)J}h^{(n/2-1)IJ} \alpha
 + \frac{1}{(n-2)^2} \varphi_{(p)J}h^{(n/2-1)IJ}\nabla^2\alpha 
 + \frac{n-3}{n-2} 
 h^{(n/2-1)IJ}\nabla_K \varphi_{(p)J}\nabla^K\alpha\Big]
\Bigg], \label{j-trans}
\end{eqnarray}
%
where we used Eq.~(\ref{j-radiation}), the Killing equation 
$\nabla_{I}\varphi_{(p)J}+\nabla_{J}\varphi_{(p)I}=0$ 
and $\nabla_{I}\nabla_{J}\alpha=\omega_{IJ}\nabla^2\alpha/(n-2)$.

Using Eq.~(\ref{j-trans}), the variation of the Bondi angular momentum $\delta J^{\text{Bondi}}_{(p)}$ becomes 
%
\begin{eqnarray}
\delta J^{\text{Bondi}}_{(p)}&=&-\frac{n-1}{16\pi G}\int_{S^{n-2}}\varphi^{I}_{(p)}\delta j_{I} d\Omega \notag \\
&=&\frac{1}{16\pi G}\int_{S^{n-2}}\alpha\varphi^{I}_{(p)} \Bigg[ 2\dot{h}^{(1)}_{IJ}\nabla_{K}h^{(1)JK}
-\nabla_{K}h_{IJ}^{(1)}\dot{h}^{(1)JK} +\frac{1}{2}\dot{h}^{(1)JK}\nabla_{I}h^{(1)}_{JK}\Bigg] d\Omega \notag \\
&&-\frac{n-2}{16\pi G}\int_{S^{n-2}}m\varphi^{I}_{(p)}\nabla_{I}\alpha d\Omega.  \label{J-trans}
\end{eqnarray} 
%
Note that the total derivative terms in Eq.~(\ref{j-trans}) do not contribute to $\delta J^{\text{Bondi}}_{(p)}$.

From the result of Eq.~(\ref{J-trans}), we can show that
Eq.~(\ref{Bondi-poincare}) holds as follows. 
We note that a rotational Killing vector $\varphi^{I}_{(p)}$ can be
rewritten as 
$\varphi^{I}_{(p)} = \varphi_{ij} \hat{x}^{(i)}\nabla^I \hat{x}^{(j)}$
where $\varphi_{ij}$ is a constant anti-symmetric tensor with
$(n-1)(n-2)/2$ independent components. 
Using the relation 
$\nabla_I \hat{x}^{(i)} \nabla^I \hat{x}^{(j)} = \delta^{ij} - \hat{x}^{(i)}\hat{x}^{(j)}$, 
we have 
$\varphi^{I}_{(p)}\nabla_{I}\alpha = \varphi_{ij}\alpha^j \hat{x}^{(i)}$.
Since $J^{\text{Bondi}}_{(p)}$ is expressed by 
$J^{\text{Bondi}}_{(p)} = \varphi^{ij}M^{\text{Bondi}}_{ij}$,
Eq.~(\ref{J-trans}) yields 
\begin{equation}
 \varphi^{ij}M^{\text{Bondi}}_{ij} \rightarrow 
  \varphi^{ij}\Big[ M^{\text{Bondi}}_{ij}
  - 2P^{\text{Bondi}}_{[i}\alpha_{j]}
  + \alpha^\mu \frac{d}{du} M^{\text{Bondi}}_{ij\mu}\Big], \label{M-trans}
\end{equation}
where we wrote  
\begin{equation}
 \varphi^{ij} \frac{d}{du} M^{\text{Bondi}}_{ij\mu}
  = \frac{1}{16\pi G}\int_{S^{n-2}}\hat{x}_\mu \varphi^{I}_{(p)} \Bigg[ 2\dot{h}^{(1)}_{IJ}\nabla_{K}h^{(1)JK}
-\nabla_{K}h_{IJ}^{(1)}\dot{h}^{(1)JK} +\frac{1}{2}\dot{h}^{(1)JK}\nabla_{I}h^{(1)}_{JK}\Bigg] d\Omega.
\end{equation}
Note that the indices $\mu$, $\nu$, $\dots$ and $i$, $j$, $\dots$ are
raised and lowered by the $n$-dimensional Minkowski metric and the
$(n-1)$-dimensional Euclidean flat metric, respectively.
Consequently, we could show that the Bondi angular momentum $J^{\text{Bondi}}_{(p)}$ 
is transformed covariantly as Eq.~(\ref{Bondi-poincare}).

Here we have a comment on the four dimensional cases. 
Because there is no condition on $\alpha$ in four dimensions, we cannot obtain the expressions 
corresponding to Eqs.~(\ref{j-trans}) and (\ref{J-trans}). Therefore the Bondi angular momentum 
is not transformed as Eq. (\ref{M-trans}) in four dimensions. In fact, the variation of the Bondi angular 
momentum has additional contributions from supertranslations, which are given by $l>1$ modes of spherical 
harmonics in $\alpha$. This is called the supertranslation ambiguity of the angular momentum at null infinity. 
Hence, we cannot have well-defined notion of the angular momentum at null infinity in four dimensions.

\section{summary and outlook}

In this paper we defined the Bondi angular momentum at null infinity 
in arbitrary higher dimensions and showed its covariant property with 
respect to the asymptotic symmetry at null infinity.  
The asymptotic symmetry becomes the Poincar\'e group in 
higher dimensions than four. This means that we can choose the $n$ 
directions of the translation without any ambiguities at null infinity. 
Then the angular momentum with the rotational axis can be defined. 
In four dimensions, on the other hand, the asymptotic symmetry at null infinities 
has the 
supertranslation, not the translation. The supertranslation has the infinite 
directions of the translation. Hence there are ambiguities of the choice of the rotational axis. 
This effects of the freedom in the definition of the rotational axis 
due to the supertranslation cannot be distinguished from the 
contributions of the variation of angular momentum by gravitational 
waves. This is the reason why we cannot define the angular momentum 
at null infinity in four dimensions. 

As one of applications of our analysis, there is the investigation of 
the peeling theorem in higher dimensions \cite{Bondi:1962px,Sachs:1962wk,Godazgar:2012zq}. The 
peeling property has played an important role in the study of the gravity 
in four dimensions, such as the stability analysis of black holes and 
construction of the exact solutions. We expect that the peeling theorem 
is useful in higher dimensions too. Using our results, general higher 
dimensional spacetimes with gravitational waves are classified by the decaying 
rate of the Weyl tensor or some geometric quantities. The effort for this 
direction has been reported \cite{Godazgar:2012zq}.

\section*{Acknowledgment}
KT is supported by JSPS Grant-in-Aid for Scientific Research (No.~21-2105).  
This work is supported in part by MEXT thorough Grant-in-Aid
for Scientific Research (A) No.~21244033 (TS) and Grant-in-Aid for
Creative Scientific Research No.~19GS0219 (TS and SK).
This work is also supported in part by MEXT through Grant-in-Aid for the
Global COE Program ``The Next Generation of Physics, Spun from
Universality and Emergence'' at Kyoto University.

\end{document}